\documentclass[aps,prl,twocolumn,amsmath]{revtex4}
\usepackage{amsmath}
\usepackage{amssymb}
\usepackage{bm}
\usepackage{epsfig}
\usepackage{graphicx}
\usepackage{color}

\begin{document}

\title{Optical measurements of spin noise as a high resolution spectroscopic tool}

\author{Valerii~S. Zapasskii$^1$, Alex Greilich$^2$, Scott~A. Crooker$^3$, Yan Li$^3$, Gleb~G. Kozlov$^1$, Dmitri~R. Yakovlev$^{2,4}$, Dirk Reuter$^5$, Andreas~D. Wieck$^5$, and Manfred Bayer$^2$ }

\affiliation{$^1$ St. Petersburg State University, Spin Optics Laboratory, 198504 St. Petersburg, Russia}

\affiliation{$^2$ Experimentelle Physik 2, Technische Universit\"at Dortmund, D-44221 Dortmund, Germany}

\affiliation{$^3$ National High Magnetic Field Laboratory, Los Alamos, NM 87545, USA}

\affiliation{$^4$ Ioffe Physical-Technical Institute, Russian Academy of Sciences, 194021 St. Petersburg, Russia}

\affiliation{$^5$ Angewandte Festk\"orperphysik, Ruhr-Universit\"at Bochum, D-44780 Bochum, Germany}

\begin{abstract}

\end{abstract}

\maketitle

\textbf{The intrinsic fluctuations of electron spins in
semiconductors and atomic vapors generate a small, randomly-varying
``spin noise" that can be detected by sensitive optical methods such
as Faraday rotation~\cite{AZ, Crooker, Mueller}. Recent studies have
demonstrated that the frequency, linewidth, and lineshape of this
spin noise directly reveals dynamical spin properties such as
dephasing times, relaxation mechanisms and g-factors \emph{without}
perturbing the spins away from
equilibrium~\cite{Oestreich:GaAs,Mueller:QW,Crooker:PRB,Yan:PRL}.
Here we demonstrate that spin noise measurements using
wavelength-tunable probe light forms the basis of a powerful and
novel \emph{spectroscopic} tool to provide unique information that
is fundamentally inaccessible via conventional linear optics. In
particular, the wavelength dependence of the detected spin noise
\emph{power} can reveal homogeneous linewidths buried within
inhomogeneously-broadened optical spectra, and can resolve
overlapping optical transitions belonging to different spin systems.
These new possibilities are explored both theoretically and via
experiments on spin systems in opposite limits of inhomogeneous
broadening (alkali atom vapors and semiconductor quantum dots).}

Optical probes of spin and magnetization dynamics enjoy broad
applications across the many atomic, semiconductor, and metallic
systems in which spin-orbit interactions allow coupling between
electron spin polarization and the optical (circular) polarization
of light~\cite{Happer,Dyak}. Experimental techniques include
powerful spectroscopies based on optical Faraday/Kerr rotation,
circular dichroism, or circularly-polarized
photoluminescence~\cite{ZK}. Using these methods, signals are
typically proportional to some intentionally-induced change in net
magnetization or spin polarization. Recently, however, sensitive
magnetometers based on optical Faraday rotation (FR) have detected
the tiny stochastic fluctuations of \emph{unperturbed} spins in
thermal equilibrium~\cite{AZ, Crooker, Mueller}. In accord with the
fluctuation-dissipation theorem, the radio-frequency spectrum of
this intrinsic ``spin noise" reveals the same dynamical spin
properties as in conventional magnetic resonance techniques, but
without driving the spin system away from equilibrium~\cite{FDT}.

The use of FR to detect spin noise is motivated by its sensitivity
to spin polarization in materials with spin-orbit coupling. This
coupling leads to the well-known optical selection rules that
preferentially couple optical transitions from spin-up or spin-down
electron states to right- or left-circularly polarized light,
respectively~\cite{OptOrient}. Examples of such systems include
band-edge transitions in III-V semiconductors, and the $S
\rightarrow P$ transitions of alkali atoms.

The energy-dependent Verdet constant $V(E)$ of a material
characterizes how much FR is elicited (per unit applied magnetic
field and per unit system length) when light of photon energy $E$ passes through
the material. $V(E)$, being sensitive to the
difference between right- and left- circular indices of refraction,
typically exhibits a dispersive lineshape near an isolated
spin-sensitive optical transition, passing through zero at line
center~\cite{caveat}. When using FR to detect spin noise, it
might therefore seem sensible to tune the probe wavelength to the
peak of $V(E)$, and avoid energies where $V(E) \rightarrow 0$.
However, owing to the correlated or uncorrelated nature of spin
fluctuations in different spin systems, we shall show that this is
not always the case, and that in general there is no direct
connection between $V(E)$ and the magnitude of the measured spin
noise.

Here we demonstrate that photon energy-dependent measurements of
optical spin noise (henceforth called `OSN spectra') can provide
highly specific information about spin-dependent optical
transitions, and can function as a novel and powerful spectroscopic
tool.

In general the optical spectrum of a paramagnetic material, whose
ground state represents the spin system of interest, is composed of
multiple bands. Each band makes a contribution $\theta_i(E)$ to the
total FR $\theta(E)$ (see Methods):

\begin{equation}
\theta(E) =\sum_i\theta_i(E) \sim \sum_i V_i(E). \label{Eq:1}
\end{equation}

Here, $V_i(E)$ is the corresponding partial Verdet constant of the
paramagnet. Note that we are considering only the
\emph{paramagnetic} contributions to $\theta_i(E)$ connected with magnetization of the spin
system~\cite{B&S,Zap}.

We focus on the quantity
$\langle\delta\theta^2(E)\rangle$, the mean-square
fluctuations of FR due to spin fluctuations. It can contain fluctuating contributions from multiple
spectral bands. When calculating its spectral dependence, however, we must distinguish
between two fundamentally different cases. In the first case, all
the constituent spectral bands are related to a single spin system,
such that \emph{FR fluctuations from different bands occur
synchronously and are correlated}. Therefore $\langle\delta\theta^2(E)\rangle$ will vary with $E$ as the
mean-square sum of all the partial contributions from which interference terms between the bands may arise:
\begin{equation}
\langle\delta\theta^2(E)\rangle \sim \langle [\sum_i
V_i(E)]^2\rangle.
\label{Eq:2}
\end{equation}

In the second case, the various spectral bands are each related to a
different spin system, meaning that spin fluctuations in
different bands are \emph{uncorrelated}. Here, $\langle\delta\theta^2(E)\rangle$ must be computed as the sum of the
mean-square partial fluctuations, so that no interference occurs:
\begin{equation}
\langle\delta\theta^2(E)\rangle \sim \sum_{i}\langle
[V_i(E)]^2\rangle.
\label{Eq:3}
\end{equation}

Crucially, this distinction is significant for closely-spaced bands, and becomes most
important when an overlapping multitude of nominally unresolvable spectral components exists
(as for inhomogeneously broadened optical spectra).

\begin{figure}
\begin{center}
\includegraphics [width=8cm]{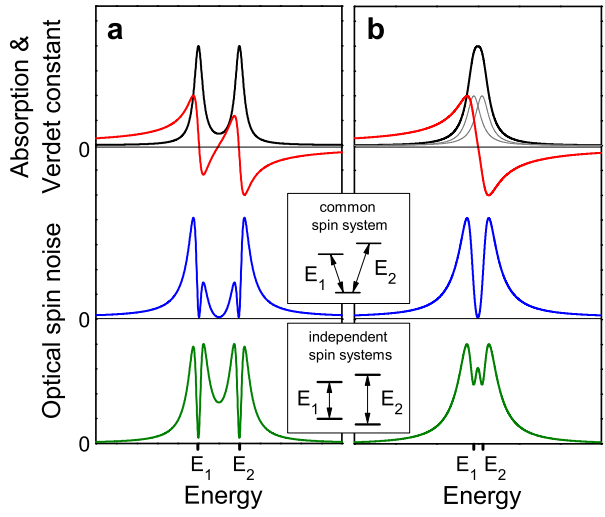}
\caption{\textbf{Optical spin noise of common and independent spin
systems.} Modeled optical spectra of a spin system having
\textbf{a}, resolved and \textbf{b}, unresolved spectral lines in
absorption. The top panels show the absorption (black) and Faraday
rotation (Verdet) spectra (red). The middle and lower panels show
the expected optical spin noise (OSN) spectra for the case where the
two optical transitions originate from a common spin system
(correlated spin fluctuations), or from independent spin systems
(uncorrelated fluctuations), respectively.}
\label{fig1}
\end{center}
\end{figure}

Figure~\ref{fig1}a demonstrates this distinction for a simple model
optical spectrum containing two well-resolved and
homogenously-broadened bands. Shown are the absorption spectrum, FR
spectrum, and associated OSN spectra for the case where the two
bands originate from a common spin system, and for the case of two
independent spin systems, calculated using Eqs.~(\ref{Eq:2}) and
(\ref{Eq:3}) respectively. Note that the conventional FR spectrum
calculated using Eq.~(\ref{Eq:1}) remains the same whether the two
optical transitions derive from a common spin system or not --
linear magneto-optics cannot distinguish between these two cases. In
marked contrast, the OSN spectra are fundamentally different:
$\langle\delta\theta^2(E)\rangle$ can be large between the two
components for the case of independent spin systems (bottom curve),
even though the net FR vanishes.

A second example of this essential distinction and of the
informative potential of OSN spectroscopy is presented in
Fig.~\ref{fig1}b, which shows absorption, FR, and OSN spectra for
two overlapping spectral bands. Clearly, the two transitions are not
resolved in absorption or in FR. However, they \emph{are} resolved
in the OSN spectrum when the bands originate from two independent
spin systems, because the spin fluctuations are uncorrelated.

\begin{figure}
\begin{center}
\includegraphics [width=12cm]{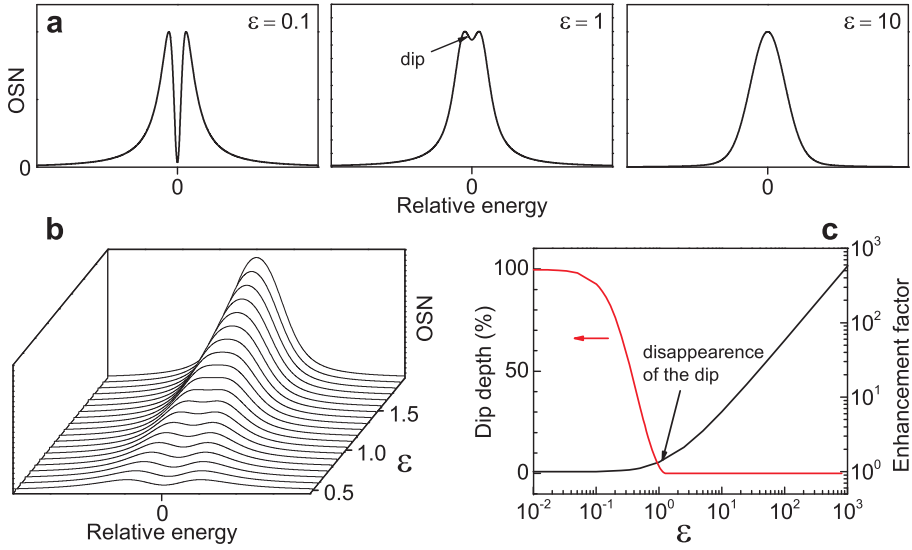}
\caption {\textbf{Effect of inhomogeneous to homogeneous linewidth
relation on optical spin noise of independent spin systems}.
\textbf{a}, Optically-detected spin noise spectra for spin-dependent
optical transitions exhibiting different ratios of inhomogeneous to
homogeneous linewidth, $\varepsilon = \gamma_{inh}/\gamma_h$ = 0.1,
1, and 10. \textbf{b}, Evolution of the OSN spectrum with increasing
$\varepsilon$. \textbf{c}, OSN enhancement factor and evolution of
the dip depth vs. $\varepsilon$.}
\label{fig2}
\end{center}
\end{figure}

These examples show that OSN spectroscopy, which uses only weak
optical fields, opens access to information lying beyond the
potential of conventional linear optics. A fascinating illustration
of this fact can be seen in Fig.~\ref{fig2}, which considers the
evolution of OSN spectra for independent spin systems as optical
transitions become increasingly dominated by inhomogeneous
broadening. In this limit the absorption, FR, and OSN spectra are
calculated as the convolution of the corresponding homogeneous
spectrum with a broader Gaussian function representing the
inhomogeneous distribution of individual optical transitions (see
Methods). Figure~\ref{fig2} shows that the shape of OSN spectra
changes dramatically as the ratio of inhomogeneous to homogeneous
linewidth, $\varepsilon = \gamma_{inh}/\gamma_{h}$, increases. In
contrast, the shape of the absorption, FR spectra and spectra of
common spin system remain practically the same for all $\varepsilon$
(not shown here). When $\varepsilon \ll 1$ and the optical
transition is mostly homogeneously broadened, the OSN spectrum
approximately follows the FR squared. In this case both situations
(common and independent spin systems) show a pronounced dip at band
center. As inhomogeneous broadening increases the dip becomes
shallower, eventually disappearing when $\varepsilon
> 1$, see Fig.~\ref{fig2}b. For bands having strong inhomogeneous
broadening ($\varepsilon \gg 1$), the OSN spectrum becomes similar
to that of the absorption spectrum (\emph{i.e.}, with a maximum at
band center), \emph{despite} the fact that the FR is zero at band
center.

Additionally, the area under the OSN spectrum increases rapidly for
higher $\varepsilon$ - that is, as the homogeneous linewidths of
individual resonances (within an inhomogeneously-broadened band)
become smaller, Fig.~\ref{fig2}b. Qualitatively, this behavior can
be explained by the fact that for smaller $\gamma_{h}$, the probe
becomes much more sensitive to those resonances at small detuning
$\Delta$, since the magnitude of the FR from each individual
resonance varies as $\Delta/(\Delta^2+\gamma_h^2)$.

Figure~\ref{fig2}c summarizes these findings and shows the evolution
of the central dip in the OSN spectrum and the development of the
enhancement factor with $\varepsilon$, calculated for an absorption
band having a constant total oscillator strength. The enhancement factor
represents the ratio between the areas under the OSN spectra
calculated for independent and common spin systems. For very small
$\varepsilon$ these areas are equal, giving no enhancement, while
with increase of $\varepsilon$ the OSN of independent spin systems
prevails.

The strong dependence of the enhancement on $\varepsilon$, as well
as the depth of the dip in the OSN spectrum, can therefore be used
for evaluating the degree of inhomogeneous broadening of optical
transitions associated with spin systems.

It is interesting to note that, from an experimental viewpoint,
systems with strong inhomogeneous broadening can be considered as
very favorable objects for OSN spectroscopy, due to the enhancement
of measured spin noise power at large $\varepsilon$.

To validate and illustrate the above considerations, we present
experimental OSN data from two spin systems that correspond to the
limits of predominantly homogeneous and inhomogeneous
broadening ($\varepsilon \ll 1$ and $\varepsilon \gg 1$,
respectively).

\begin{figure}
\begin{center}
\includegraphics [width=8cm]{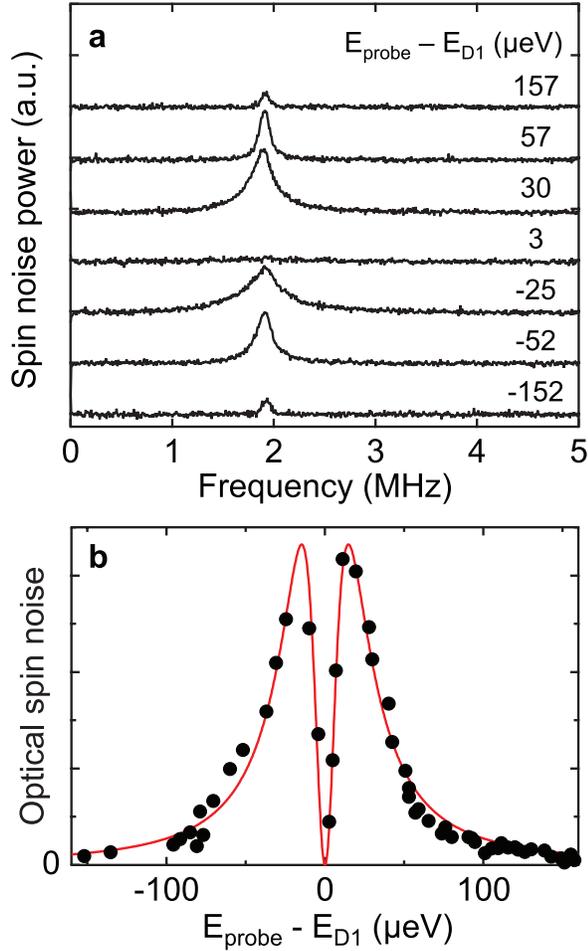}
\caption {\textbf{Optical spin noise from a homogeneously-broadened
system: D1 line of $^{41}$K vapor.} \textbf{a}, The raw spin noise
data (power spectral density) from spin fluctuations of the $4S$
electrons in $^{41}$K, acquired as the photon energy of the probe
laser was tuned through the D1 resonance at 1.60995868\,eV.
$T=110^{\circ}$C ($\sim$25\% of the probe light is absorbed on
resonance). Note that essentially no noise is observed on resonance,
as expected for a homogeneously-broadened line. A small transverse
magnetic field was applied ($\sim$3\,Gauss) to shift the spin noise
to finite frequency. \textbf{b}, OSN spectrum: the total measured
spin noise power as a function of the probe photon energy. The red
solid line is the fitting using a single squared FR spectra $\theta^2_i(E)$ with $\gamma_h = 14.74\,\mu$eV (see Methods).}
\label{fig3}
\end{center}
\end{figure}

For the case of predominantly homogeneous broadening ($\varepsilon
\ll 1$), we studied a warm (110\,$^{\circ}$C) vapor of the potassium
isotope $^{41}$K in nitrogen buffer gas (see Methods).
Figure~\ref{fig3}a shows how the raw spin noise data vary as the
probe laser is tuned through the fundamental spin-sensitive D1
transition ($4S_{1/2} \rightarrow 4P_{1/2}$). Integrating the total
noise power under these data gives the associated OSN spectrum shown
in Fig.~\ref{fig3}b. As expected (see Fig.~\ref{fig2}a,
$\varepsilon=0.1$), it is comprised of two bumps with a
well-pronounced dip at line center. This finding is supported by
recent observations on $^{87}$Rb vapor~\cite{Horn:PRA}.

To investigate OSN in the opposite regime of predominantly
\emph{inhomogeneous} broadening, we measured the spin noise from an
ensemble of hole-doped (In,Ga)As/GaAs quantum dots (QDs). At low temperatures, individual QDs exhibit
sharp atomic-like optical transitions as narrow as tens of $\mu$eV;
however, the distribution of QD sizes and compositions leads, in ensemble measurements, to a broad $\sim$20\,meV
inhomogeneous linewidth. The optical transition from a
single resident hole to a positively charged trion is right- or left
circularly polarized depending on the hole's initial spin orientation.
Hole spin fluctuations therefore generate a measurable
spin noise~\cite{Crooker:2010,Yan:PRL}.

\begin{figure}
\begin{center}
\includegraphics [width=8cm]{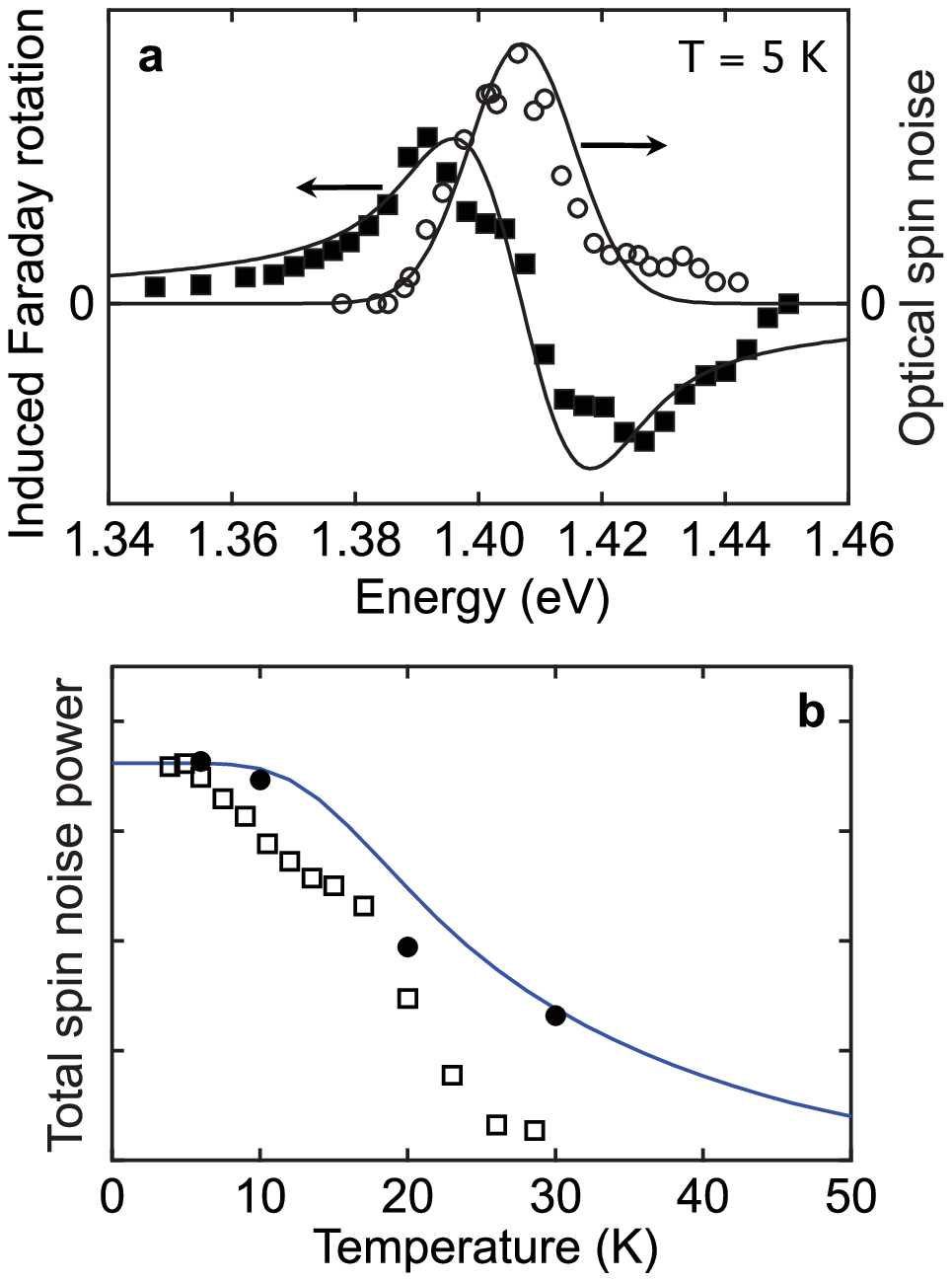}
\caption{\textbf{Optical spin noise from an
inhomogeneously-broadened system: ensemble of (In,Ga)As quantum dots
containing single hole spins.} \textbf{a}, Optical spectra of the
(intentionally) induced Faraday rotation, and of the detected total spin
noise power, as a function of probe photon energy. Solid lines are
the fits using $\theta_i(E)$ (for FR) and $\theta^2_i(E)$ (for OSN) convoluted with Gaussian representing an inhomogeneous broadening of $\gamma_{inh}=8.5$\,meV (see Methods). (As long as the
homogeneous linewidth is much narrower than the inhomogeneous
broadening, it has negligible effect on the final result.)
\textbf{b}, Temperature dependence of the total spin noise power measured at the maximum of absorption band (1.406\,eV), which
decreases with temperature as the homogeneous linewidth $\gamma_h$
of the individual quantum dots grows. Squares and points represent
two sets of data taken on the same sample at different points. Solid
line is a plot using the modeling procedure presented in the Methods
section.}
\label{fig4}
\end{center}
\end{figure}

Figure~\ref{fig4}a shows the OSN spectrum from these holes as the
probe laser was tuned through the absorption band of the QD
ensemble.  Also shown for reference is the usual FR spectrum of the
QD ensemble, obtained by intentionally polarizing the hole spins
with circularly polarized light at 1.579\,eV, and simultaneously
detecting the induced FR as a function of probe photon energy. As
predicted for the case of an inhomogeneously-broadened line (see
Fig.~\ref{fig2}a, $\varepsilon=10$), the OSN spectrum does
\emph{not} exhibit any dip at line center, but rather achieves a
maximum, even though the induced FR passes through zero.

Interestingly, additional confirmation of the model is obtained from
the temperature dependence of the total spin noise from these QDs.
As predicted above in Fig.~\ref{fig2}c, the detected spin noise from
an inhomogeneously-broadened system will change if the ratio
$\gamma_{inh}/\gamma_{h}$ varies. In these (In,Ga)As QDs,
$\gamma_{h}$ is expected to be strongly temperature dependent, even
at low temperatures~\cite{Borri}. This change is clearly apparent in
the noise measurements: Fig.~\ref{fig4}b shows the total spin noise
power measured from 4 to 30\,K, using a fixed probe energy
(1.406\,eV) corresponding to the maximum of the OSN spectrum. The
total noise decreases as $\gamma_{h}$ of the underlying QDs grows
with increasing temperature, approximately in accord with that
measured by sophisticated four-wave mixing techniques in similar
(In,Ga)As QDs~\cite{Borri}. Thus, spin noise measurements alone
provide a powerful experimental tool for penetrating the internal
structure of inhomogeneously broadened systems and for revealing
variations of homogeneous linewidths that may otherwise be obscured
by strong inhomogeneous broadening. It is especially noteworthy that
this sensitivity to the hidden fine structure of an inhomogeneously
broadened band works even when $\gamma_{inh} \gg \gamma_{h}$, making
it possible to realize an effectively high (sub-linewidth) spectral
resolution by means of a linear optical technique.

In summary, we have shown that an optical spectroscopy based on
measuring spin fluctuations can provide important information about
spin systems that is generally regarded as inaccessible for
conventional linear optical spectroscopic techniques. The
capabilities of this technique are illustrated by experimental OSN
spectra of homogeneously and inhomogeneously broadened spin systems
-- electrons in the $4S$ ground state of $^{41}$K, and resident holes in
(In,Ga)As quantum dot ensembles, respectively. These unique properties of OSN spectroscopy that make it possible to
penetrate inside the optical bands while remaining within the bounds of linear optical polarizability make it a powerful method of
research applicable to many materials important for present-day photonics and information science.

\section{Methods}
\textbf{Optical spin noise measurements.} Spin noise measurements
were performed using a weak, linearly-polarized beam from a
continuous-wave Ti:sapphire ring laser. The intrinsic spin
fluctuations of the medium along the laser direction imparted
Faraday rotation fluctuations on the transmitted laser, which were
detected by a balanced photoreceiver. The output voltage was
amplified, digitized, and fast-Fourier-transformed in real time to
obtain the power density of the measured spin noise (in units of
nanoradians$^2$/Hz), typically from 0\,Hz up to a few MHz for atoms
and up to a hundreds of MHz for QDs. A full description of the
experimental setup can be found
in~\cite{Crooker,Mueller,Crooker:2010}. To obtain the OSN spectrum,
the measured spin noise was integrated over frequency (total spin noise power) and then
recorded as a function of the photon energy of the probe beam.

\textbf{$^{41}$K vapor.} We used $^{41}$K in a 1\,cm thick glass
cell containing a significant 200\,Torr background of nitrogen
buffer gas. Collisional broadening of the fundamental $4S_{1/2}
\rightarrow 4P_{1/2}$ (D1 line) and $4S_{1/2} \rightarrow 4P_{3/2}$
(D2 line) optical transitions increased their linewidths to of order
10\,GHz, which greatly exceeds their underlying hyperfine structure
($\sim$254\,MHz) or Doppler width. In this regime these transitions
can be considered homogeneously broadened. Spin fluctuations of the
$4S$ electrons generate measurable spin noise when the probe laser
is tuned near the D1 or D2 resonance.

\textbf{(In,Ga)As QD sample.} We used a 20-layer structure of
MBE-grown (In,Ga)As QDs separated by 60\,nm GaAs barriers with the
QD density of 10$^{10}$\,cm$^{-2}$ per layer. Thermal annealing for
30\,s at 960$^{\circ}$C shifted the emission to the sensitivity
range of silicon photodetectors. The sample had a low level of
background p-type doping. It was mounted on the cold finger of an
optical cryostat and demonstrated the FR-detected spin noise
provided by resident holes~\cite{Crooker:2010}.

\textbf{Simulations.} To simulate the single band contribution we
use the following functions: $f_i(E)=
\gamma_h/((E-E_i)^2+\gamma_h^2)$ for absorption and $\theta_i(E)=
-(E-E_i)/((E-E_i)^2+\gamma_h^2)$ for FR (Verdet constant), with
$E_i$ the resonance energy and $\gamma_h$ the homogeneous linewidth.
For inhomogeneous spectra we convolute the single contribution
numerically with a Gaussian having an inhomogeneous linewidth
$\gamma_{inh}$. So, for OSN spectra in the case of independent spin
systems we obtain:
\begin{equation}\label{Eq:conv}
    \langle\delta\theta^2(E)\rangle=\frac{1}{\sqrt{2 \pi \gamma_{inh}^2} }\int
    \theta_i^2(E)e^{-E_i^2/(2\gamma_{inh}^2)}d E_i.
\end{equation}

Using a temperature dependent $\gamma_h$ in Eq.~(\ref{Eq:conv}) we
calculate the temperature dependence of the total spin noise power at single energy position. Here we use Ref.~[\onlinecite{Borri}], where the very similar QD structures
were studied to define the functional dependence of homogeneous
linewidth on temperature. The QD sample provides the confinement
energy of $E_c=89$\,meV, defined by energy distance between the QD
emission from the ground state and wetting layer. It leads us to the
following parameters: $\gamma_0=2.6\,\mu$eV, $b_1=30\,\mu$eV and
$b_2=4.2$\,meV in the equation:
\begin{equation}\label{Eq:4}
\gamma_h(T)=\gamma_0+\frac{b_1}{e^{E_1/k_BT}-1}+\frac{b_2}{e^{E_2/k_BT}-1},
\end{equation}
where $E_1=6$\,meV, and $E_2=28$\,meV.

\newpage

\begin{thebibliography}{99}

\bibitem{AZ} Aleksandrov, E. B. and Zapasskii, V. S. Magnetic resonance in the Faraday-rotation noise
spectrum. \textit{Zh. Eksp. Teor. Fiz.} \textbf{81,} 132--138 (1981)
[\textit{JETP} \textbf{54,} 64--67 (1981)].

\bibitem{Crooker} Crooker, S. A., Rickel, D. G., Balatsky, A. V. and Smith, D. Spectroscopy of spontaneous spin
noise as a probe of spin dynamics and magnetic resonance.
\textit{Nature} {\bf 431,} 49--52 (2004).

\bibitem{Mueller} M\"uller, G. M., Oestreich, M., R\"omer, M. and Hubner, J. Semiconductor Spin Noise Spectroscopy:
Fundamentals, Accomplishments, and Challenges. \textit{Physica E}
{\bf 43,} 569--587 (2010).

\bibitem{Oestreich:GaAs} Oestreich, M., R\"omer, M., Haug, R. J. and H\"agele, D. Spin Noise Spectroscopy in GaAs. \textit{Phys. Rev. Lett.} \textbf{95,} 216603 (2005).

\bibitem{Mueller:QW} M\"uller, G. M. \textit{et al.} Spin Noise Spectroscopy in GaAs (110) Quantum Wells: Access to Intrinsic Spin Lifetimes and Equilibrium Electron Dynamics. \textit{Phys. Rev. Lett.} \textbf{101,} 206601 (2008).

\bibitem{Crooker:PRB} Crooker, S. A., Cheng, L. and Smith, D. L. Spin noise of conduction electrons in $n$-type bulk GaAs. \textit{Phys. Rev. B} \textbf{79,} 035208 (2009).

\bibitem{Yan:PRL} Li, Yan \textit{et al.} Intrinsic Spin Fluctuations Reveal the Dynamical Response Function of Holes Coupled to Nuclear Spin Baths in (In,Ga)As Quantum Dots. \textit{Phys. Rev. Lett.} \textbf{108,} 186603 (2012).

\bibitem{Happer} Happer, W. Optical Pumping. {\it Rev. Mod. Phys.} {\bf 44,} 169--249
(1972).

\bibitem{Dyak} \textit{Spin Physics in Semiconductors}, edited by Dyakonov, M. I.
(Springer-Verlag, Berlin, 2008).

\bibitem{ZK} Zvezdin, A. K. and Kotov, V. A. \textit{Modern Magnetooptics and Magnetooptical Materials} (Institute of Physics, Bristol,
1997).

\bibitem{FDT} Kubo, R. The fluctuation-dissipation theorem. \textit{Rep. Prog. Phys.} \textbf{29,} 255-–284 (1966).

\bibitem{OptOrient} Dyakonov, M. I., Perel, V. I. \textit{Optical Orientation} Ch. 2, edited by Meier, F., Zakharchenya, B. P. (North-Holland, Amsterdam,
1984).

\bibitem{caveat} For small magnetic (Zeeman) splittings less than the linewidth of the optical transition.

\bibitem{B&S} Buckingham, A. D. and Stephens, P. J. Magnetic Optical Activity. \textit{Ann. Rev. Phys. Chem.} {\bf 17,}
399--432 (1966).

\bibitem{Zap} Zapasskii, V. S. {\it Spectroscopy of Solids Containing Rare-Earth Ions} p. 674, Eds. Kaplyanskii A. A. and Macfarlane, M. F. (Elsevier Science Publishers B.V., 1987).

\bibitem{Horn:PRA} Horn, H. \textit{et al.} Spin-noise spectroscopy under resonant optical probing conditions: Coherent and nonlinear effects. \textit{Phys. Rev. A} \textbf{84,} 043851 (2011).

\bibitem{Crooker:2010} Crooker, S. A. \textit{et al.} Spin Noise of Electrons and
Holes in Self-Assembled Quantum Dots. \textit{Phys. Rev. Lett.} {\bf 104,} 036601 (2010).

\bibitem{Borri} Borri, P., Langbein, W. and Woggon, U. Exciton dephasing via phonon interactions in InAs quantum dots: Dependence on quantum confinement. \textit{Phys. Rev. B} {\bf 71,} 115328 (2005).


\end{thebibliography}
\end{document}